# Electronic and energetic properties of Ge(110) pentagons

Pantelis Bampoulis, Adil Acun, Lijie Zhang, and Harold J.W. Zandvliet

Physics of Interfaces and Nanomaterials, MESA+ Institute for Nanotechnology, University of Twente, P.O. Box 217, 7500AE Enschede, The Netherlands

**The electronic and energetic properties of the elementary building block, i.e. a five-membered atom ring (pentagon), of the Ge(110) surface was studied by scanning tunneling microscopy and spectroscopy at room temperature. The Ge(110) surface is composed of three types of domains: two ordered domains ((16x2) and c(8x10)) and a disordered domain. The elementary building block of all three domains is a pentagon. Scanning tunneling spectra recorded on the (16x2), c(8x10) and disordered domains are very similar and reveal three well-defined electronic states. Two electronic states are located 1.1 eV and 0.3 eV below the Fermi level respectively, whereas the third electronic state is located 0.4 eV above the Fermi level. The electronic states at -0.3 eV and 0.4 eV can be ascribed to the pentagons, whilst we tentatively assigned the electronic state at -1.1 eV to a Ge-Ge back bond or trough state. In addition, we have analyzed the straight [1-12] oriented step edges. From the kink density and kink-kink distance distributions we extracted the nearest neighbor interaction energy between the pentagons, which exhibit a strong preference to occur in twins, as well as the strain relaxation energy along the pentagon-twin chains.**

**Introduction**

The low-index surfaces of the group IV semiconductors have been studied in great detail, except the (110) surface [1-14]. This is quite remarkable given the fact that the (110) is intrinsically anisotropic, in contrast to its (100) and (111) counterparts. The surface free energy per unit area of the (110) surface is higher than that of the (100) and (111) surfaces and therefore the (110) surface has the tendency to facet. The bulk truncated Ge(110) surface has a rectangular symmetry and is composed of zigzag rows of atoms that run in the [1-10] direction. The Ge(110) surface reconstructs into rather large unit cells, which are very complex and involve several atomic layers. The most common surface reconstructions of Ge(110) are the (16x2) and c(8x10) reconstructions. Despite a number of detailed studies there is no full consensus yet on the exact model for these reconstructions [1-12]. The (16x2) and c(8x10) are both composed of five-membered atom rings, hereafter referred as pentagons and feature small (17 15 1) facets at the steps. In the remainder of our paper we will adapt the structural models that have been put forward by Ichikawa [8,9]. These structural models of the (16x2) and c(8x10) are consistent with existing scanning tunneling microscopy data. A careful reflection high energy electron diffraction study by Ichikawa, Fujii and Sugimoto [13] revealed that prolonged annealing at temperatures below 650 K resulted in a (16x2) reconstructed surface. Therefore we can safely conclude that the (16x2) reconstruction is the thermodynamically most stable reconstruction, whereas the c(8x10) reconstruction is only a metastable and transient reconstruction. The (16x2) reconstruction undergoes an order-disorder transition at temperature of about 700 K. For a detailed description and discussion of the (16x2) and c(8x10) reconstructions, as well as a brief overview of the history of the Ge(110) surface we refer to Ichikawa's papers [2,3,8-10,13] as well as a recent paper by Mullet and Chiang [14]. The vast majority of papers published on Ge(110) deal with the structural properties of the surface, whereas the electronic properties received much less attention [15-17]. Here we present a combined scanning tunneling microscopy and spectroscopy study of the elementary building block, i.e. a pentagon, of the Ge(110) surface. The local density of states of the two most common reconstructions of the Ge(110) surface, i.e. the (16x2) and c(8x10) reconstructions, will be extracted from spatially resolved scanning tunneling spectra. The energetic interaction between the pentagons and the long-range strain relaxation within in the zigzag pentagon rows will be extracted from a statistical analysis of the roughness of the [1-12] oriented steps and the kink-kink length distribution within the steps, respectively.

**Experimental**

The scanning tunneling microscopy (STM) and spectroscopy (STS) measurements were performed at room temperature in an ultra-high vacuum system with a base pressure of $3.10^{-11}$ mbar. The Ge(110) samples were cut from nominal flat, single-side polished nearly intrinsic (50-60 $\Omega$cm) *n*-type wafers. After cutting, the samples were thoroughly cleaned with isopropanol alcohol before inserting them into the ultra-high vacuum system. Firstly, the Ge(110) samples were outgassed for at least 12 hours at a temperature of 750-800 K. Secondly, we cleaned the samples by a method that we applied to and tested extensively on the closely related Ge(001) surface [18]. This cleaning method involves several cycles of Argon ion bombardment followed by annealing at temperatures of 1100 (±25) K. After five to seven of these cleaning cycles the Ge(110) samples were atomically clean and exhibited well-ordered reconstructed (16x2) and c(8x10) domains as well as some disordered regions. The relative occupation of the various reconstructions can be tuned by varying the cooling time after a high temperature anneal. A slow cooling rate leads to an increase of the (16x2) domains at the expense of the c(8x10) and disordered domains. Since we aim at a detailed study of all domains we have rapidly cooled down our samples.

**Results and Discussion**

In Figure 1A a large-scale filled-state STM image of a Ge(110) surface is shown. The Ge(110) surface exhibits regions with c(8x10) and (16x2) reconstructions, as well as some disordered regions. In the top right quadrant of Fig. 1A a few (8x2) domains can be seen. These (8x2) have been reported earlier in ref. [14]. In the insets of Figure 1 zoom-ins of the different phases are displayed. Interestingly, all phases, including the disordered phase, are composed of five-membered atom rings (hereafter referred as pentagons). In Figures 2A and 2B simplified structural models of the (16x2) and c(8x10) reconstructions are shown. For the fully relaxed models we refer to the work of Ichikawa [8,9]. In these papers fully relaxed models of the (16x2) and the c(8x10) reconstructions are shown that are consistent with existing STM data. We would like to emphasize that the (16x2) reconstruction, in contrast to the c(8x10) reconstruction, consists of alternating up and down row of pentagons (see Fig. 2 and Figs. 1C-D). The zigzag rows of pentagons in the c(8x10) domains are aligned along the [2-25] direction, whilst the zigzag rows of pentagons in the (16x2) domains are aligned along the [1-12] direction. Both straight and rough step edges are found on the surface. The straight steps are aligned along the zigzag pentagon rows of the (16x2) domains, i.e. along the [1-12] direction. Both zigzag rows seem to be composed of pentagon twins, however a more careful inspection of the models (see refs. [8,9]) reveals that the pentagon twins are not exactly the same. For the remainder of this work we will focus our attention on the pentagon twins of the most stable reconstruction, i.e. the (16x2) reconstruction.

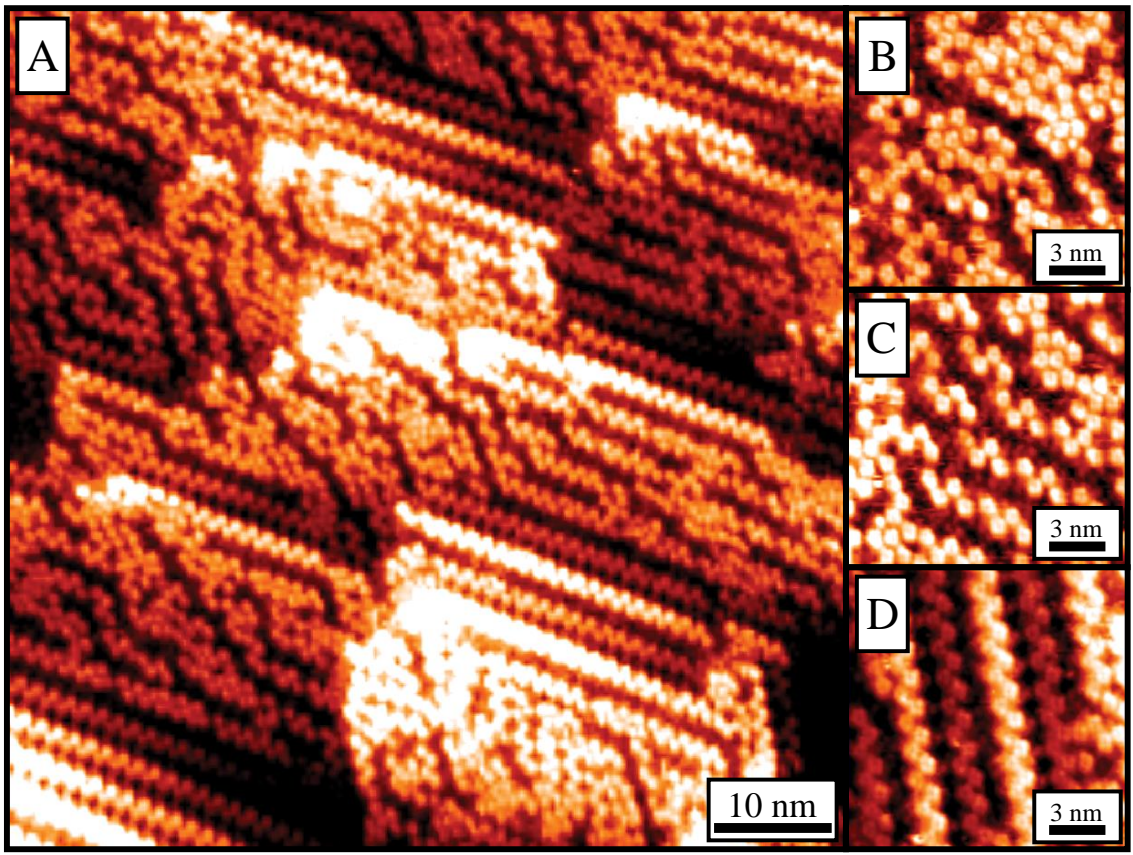

**Figure 1** (A) Filled-state scanning tunneling microscopy image of Ge(110). (B) Scanning tunneling microscope image of the disordered phase. (C) Scanning tunneling microscope image showing a region that exhibits a c(8x10) phase (middle part of the image) as well as a disordered phase (at the left and right border of the image). (D) Scanning tunneling microscope image of the (16x2) phase. Set points are in (A) -1.5 V, 0.5 nA, (B) -1.5 V, 0.29 nA, (C) -1.5 V, 0.29 nA and (D) -1.5 V and 0.29 nA.

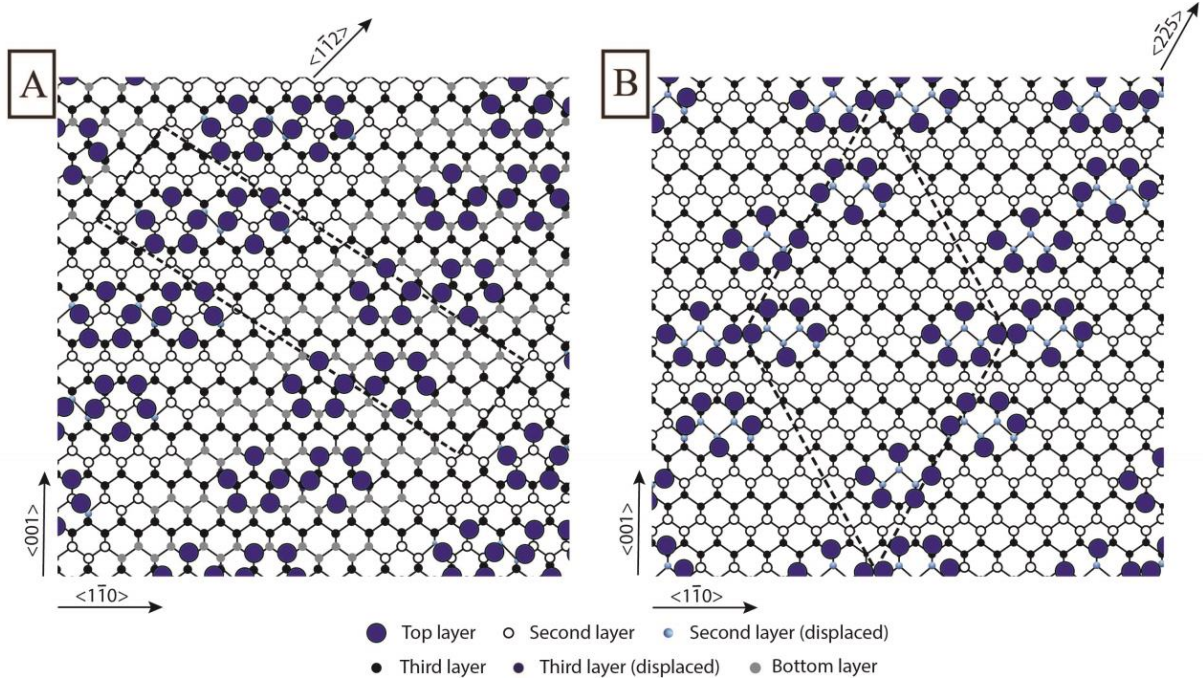

**Figure 2**

Simplified ball-and-stick models of the (16x2) reconstruction (A) and the c(8x10) reconstruction (B). Fully relaxed models for both reconstructions can be found in refs. [8,9].

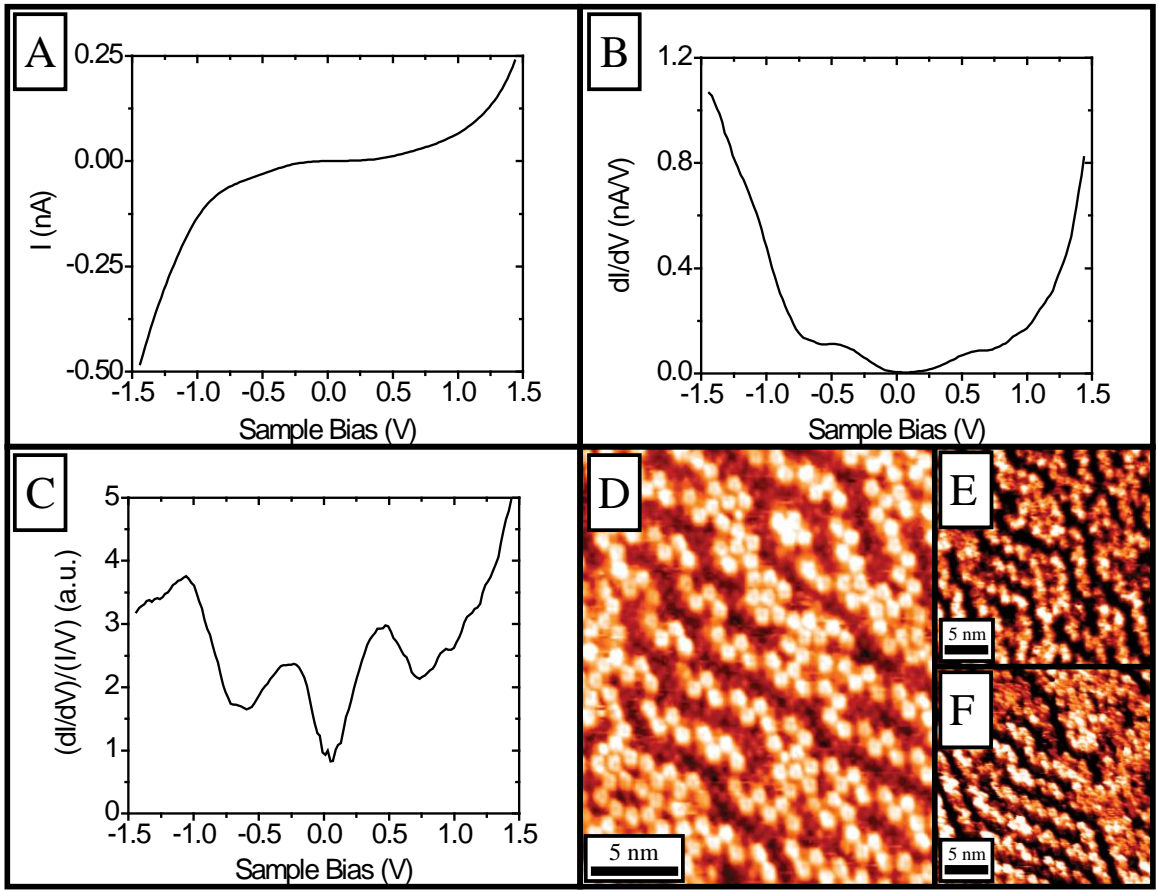

**Figure 3** (A) Current-voltage (*IV*) spectrum of the Ge(110) surface. (B) Differential conductivity *(dI/dV)* versus sample bias of the Ge(110) surface. (B) Normalized differential conductivity *(dI/dV)/(I/V)* versus sample bias of the Ge(110) surface. Set points for (A),(B) and (C) are -1.5 V, 0.25 nA. (D) Scanning tunneling microscope image of the Ge(110) surface at -1.5 V and 0.29 nA. (E) Scanning tunneling microscope image of the Ge(110) surface taken at +0.4 V and 0.15 nA. Scanning tunneling microscope image of the Ge(110) surface taken at -0.35 V and 0.15 nA .

In Figure 3A the *IV* spectrum of the bare Ge(110) surface is depicted. The *IV* curves recorded on the different regions of the Ge(110) surface, i.e. the (16x2), c(8x10) and disordered domains, are essentially the same. This is not so strange since all reconstructions/phases are composed of the same elementary building blocks, i.e. the pentagons. In addition, IV curves recorded on the pentagons of the different regions (domains) do not reveal any detectable differences.

In Figure 3B and 3C the differential conductivity *(dI/dV)* and the normalized differential conductivity *((dI/dV)/(I/V))* are shown. All IV traces are recorded with the same set points of V=-1.5 V and I=0.25 nA. In Figure 3D an STM image recorded at the used set points is depicted. In the normalized differential conductivity spectrum three well-defined peaks are resolved: two filled-states at -1.1 V and -0.3 V respectively and an empty-state at 0.4 V. In Figures 3E and 3F STM images recorded at 0.4 V and -0.35 V are shown. In contrast to figure 3D, where a clear structure in between the zigzag pentagon rows appears, figures 3E and 3F are composed of pentagons only. Since to the best of our knowledge no spectroscopic data of the Ge(110) is available, we compare our results with spectroscopic data recorded on the closely related Si(110)-(16x2) surface [19-21]. Setvín et al. [19] performed a very detailed study on the electronic structure of the Si(110) surface. In contrast to the Ge(110) surface, Setvín et al. found at least three electronic states for the Si(110) surface. Two electronic states are located very close to the Fermi level, one about 0.2 eV below the Fermi level and one about 0.2 eV above the Fermi level. These two states near the Fermi level can be ascribed to the pentagons, which are also the elementary building block of the Si(110)-(16x2) surface. Another filled state at -1.5 V is located in the middle of the pentagons and was ascribed to the underlying zigzag line of Si atoms. The energy of this electronic state is high enough to be assigned to Si-Si back bonds. The electronic states we have found for Ge(110) at -0.3 V and 0.4 V are both related to the pentagons, but our room temperature STS data has insufficient spatial resolution to resolve the exact position of the electronic states within the pentagons. The electronic state of Ge(110) located at -1.1 V is also clearly present in the troughs between the zigzag pentagon rows and therefore we tentatively ascribe this state to Ge-Ge back bonds or edge states located in the trough between the pentagon rows. The electronic state at -1.1 eV is in excellent agreement with photoemission data reported by Santoni et al. [15]. For Si(110)-(16x2) two more empty states are reported [16], one located at 1.2 V and another located at about 1.6-2.0 eV above the Fermi level. Since we have not found any empty states up to 1.5 eV for Ge(110) we will not further elaborate on these empty states.

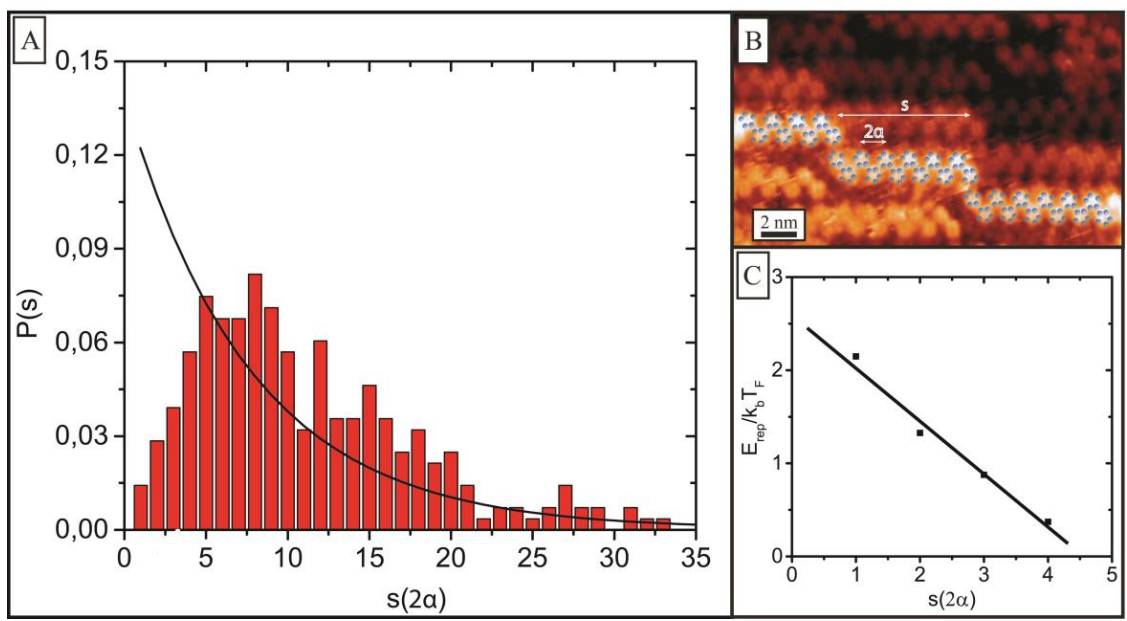

**Figure 4** (A) Histogram of the probability that two adjacent kinks are separated by a distance s (in= units 2α). The solid line refers to the independent kink model distribution, $P(s)=P_k(1-P_k)^{s-1}$, where $P_k$ is the probability of finding a kink of any kind in the step edge. (B) Scanning tunneling microscopy image of Ge(110) recorded at -1.5V and 0.29 nA showing a step edge which is aligned along the zigzag rows of the (16x2) reconstruction, i.e. the [1-12] direction. The step edge exhibits two kinks that are separated by a distance s. (C) Repulsive energy between kinks in units of $k_bT_F$ versus s. The solid line is a guide to the eye.

Subsequently we have performed a statistical analysis of the step edge roughness of steps running along the zigzag pentagon rows of the (16x2) reconstruction. The straight parts of the step edges run in the [1-12] direction. Due to thermal excitation the step edges exhibit kinks, which make the step wander. Despite the fact that the creation of a kink costs energy, there is also a substantial gain in entropy because kinks can be created at many locations within the step. The total number of thermally generated kinks and their distribution is adjusted such that the total free energy of the step edge is minimized [22,23]. Kinks can be discriminated in kinks that point towards the upper terrace (positive (+) kinks) or kinks that point towards the lower terrace (negative (-) kinks). We would like to emphasize that have not have taken into account the alternating up and down registry of the (16x2) pentagon rows. The distance between the centers of adjacent pentagon twins, $2\alpha$, is 1.34 nm (see Figure 4B). From the density of single kinks the nearest neighbor interaction energy, $E_{NN}$, between the elementary building blocks of the (16x2) reconstruction, i.e. the pentagon twin, can be extracted [24, 25],

$$\frac{n_{+1}n_{-1}}{n_0^2} = e^{-E_{NN}/k_b T_F} \qquad (1)$$

where $n_0$, $n_{+1}$ and $n_{-1}$ are probabilities of finding no kink, a single positive '+' kink and a single negative '−' kink, respectively. The densities of positive and negative kinks are very comparable and therefore the nominal miscut of the step edges with respect to the high symmetry direction ([1-12]) is rather small. $T_F$ is the freeze-in temperature and $k_b$ Boltzmann's constant. In total we analyzed a step edge length of 4.3 μm. We obtained the following values $N_0$= 2797, $N_{+1}$=178 and $N_{-1}$=211 resulting into $E_{NN}/k_b T_F$ = 5.3±0.2 ($N_{0,+1,-1}$ refer to the total number of no, + and − kinks, respectively). Double and triple kinks do occur, but they are very rare. We would like to emphasize here that the presence of these higher order kinks does not affect the validity of Eq. (1). In order to determine the nearest neighbor interaction energy between the pentagon twins we need to know the exact freeze-in temperature of the step edge roughness. Since the (16x2) reconstruction de-reconstructs at a temperature of 700 K, this temperature can be considered as an upper bound of the freeze-in temperature. Using this upper bound we find $E_{NN}$= 315±15 meV.

In Figure 4A a histogram of the kink-kink separation probability distribution is shown. The solid line is the kink-kink separation probability distribution for the independent kink model (no kink-kink interactions). It is important to mention here that it is in principle possible to find kink-kink separations of s/2, 3s/2,… etc., however pentagons have a strong tendency to occur in twins and therefore kink-kink separations of s/2, 3s/2, .. are very rare. In the independent kink model the probability of finding a kink-kink separation s (see Figure 4B) is given by,

$$P(s) = P_k (1 - P_k)^{s-1} \qquad (2)$$

where $P_k$ is the probability of finding a kink of any kind and $(1-P_k)$ is the probability of finding no kink. The distribution function eq. (2) is normalized, i.e. $\sum_s P(s)=1$. From Figure 4A it is immediately clear that there is a strong kink-kink repulsion. This kink-kink repulsion falls off over a length scale of about 4-5 nm. In Figure 4C $-\ln\left(\frac{P_{\exp}(s)}{P(s)}\right)$ $(=E_{rep}/k_bT_F)$ is plotted versus s. This plot gives the repulsion energy (measured in units of $k_bT_F$) as a function of *s*. The repulsive kink-kink interaction falls from 2.15 $k_bT_F$ (s=1), 1.3 $k_bT_F$ (s=2), 0.85 $k_bT_F$ (s=3) to 0.4 $k_bT_F$ (s=4).

In summary, we have used scanning tunneling microscopy and spectroscopy to study the structural and electronic properties of the Ge(110) surface. We observed the coexistence of three different phases at room temperature: the (16x2) phase, the c(8x10) phase and a disordered phase. The elementary building block of three phases is a five-membered atom ring (pentagon). Scanning tunneling spectra recorded at the different phases reveal that there are hardly any differences between the phases. The scanning tunneling spectra exhibit three well-defined electronic states: two filled-states located 1.1 eV and 0.3 eV below the Fermi level and an empty state which is positioned 0.4 eV above the Fermi level. The electronic states at -0.3 eV and 0.4 eV can be ascribed to the pentagons, whilst we tentatively ascribe the electronic state located at -1.1 eV to Ge-Ge back bonds. We have also analyzed the roughness of the [1-12] oriented steps of the Ge(110) surface in order to extract the energetic coupling between the pentagon twins as well as the kink-kink interaction. The interaction energy between adjacent pentagon twins is 315±15 meV and the repulsive kink-kink interaction falls from 130 meV (s=1), 80 meV (s=2), 50 meV (s=3) to 25 meV (s=4).

**Acknowledgments**

P.B. and A.A. would like to thank the Dutch Organization for Research (NWO) for financial support. L.Z. acknowledges the China Scholarship Council (CSC) for financial support.